\documentclass{article}
\usepackage{amsfonts,amssymb, amsmath}
\usepackage[english]{babel}

\def\bq{ \begin{equation}}
\def\eq{ \end{equation}}
\def\ben{ \begin{eqnarray}}
\def\en{ \end{eqnarray}}
\def\g{{\gamma}}

\newtheorem{prop}{Proposition}
\begin{document}
\title{Hamiltonization and separation of variables for Chaplygin ball on a rotating plane }
\author{A.V. Tsiganov \\
\it\small St. Petersburg State University, St. Petersburg, Russia\\
\it\small Udmurt State University,
ul. Universitetskaya 1, Izhevsk,  Russia\\
\it\small e--mail:  andrey.tsiganov@gmail.com}

\date{}
\maketitle

\begin{abstract}
 We discuss a non-Hamiltonian vector field  appearing in consideration of a partial motion of the Chaplygin ball rolling on a horizontal plane which rotates with constant angular velocity.   In two partial cases  this vector field is expressed via Hamiltonian vector fields using a non-algebraic deformation of the canonical  Poisson bivector on $e^*(3)$.  For the symmetric ball we also calculate variables of separation, compatible Poisson brackets,  algebra of Haantjes  operators and  $2\times2$ Lax matrices.
 \end{abstract}

\section{Introduction}
\setcounter{equation}{0}
Theory of integrable systems appeared as a family of  mathematical methods which can be applied to find exact solutions of dynamical systems. The main motivation was  to determine the scope of mathematical models of real physical processes.  Explicit solutions allow us to test analytical and numerical schemes  applied to a given  mathematical model and to choose a reasonable approximation to solutions of the model.

In classical mechanics change of time is a standard tool for construction of explicit solutions of equations of motion according to Kepler \cite{kep12}, Jacobi, Maupertuis \cite{mau44}, Weierstrass \cite{w},  see discussion in  \cite{ts00,ts01}  and references within. In nonholonomic mechanics Appel \cite{app} and Chaplygin \cite{ch03,ch11} also used change of time for integrating certain  nonholonomic systems with two degrees of freedom. In modern nonholomic mechanics we have many equations of motion taking a Hamiltonian form after  suitable  symmetry reduction and time reparametrisation, see \cite{bal12,bbm15,bbm16,bl09a,bol15,bm14,guh11,ko04,fasso18,fed06,bl09,gar18,bl11} and references within. Unfortunately, in most of these publications authors discuss only the form of equations of motion instead  of exact solutions of these equations.

The main aim of this note is to compare Hamiltonization and the modern separation of variables method embedding elements of artificial intelligence such as machine learning and deep learning. Harnessing of modern computational abilities for studying  integrable  systems is naturally placed as a prominent avenue in contemporary classical and quantum mechanics.  For instance, this allows us to automate  validation of mathematical models of real physical process, see one of the collection of papers about machine learning in physics \cite{col18}. In classical  mechanics computer modeling currently consists not only of approximate numerical calculations and visualization, but also of algorithmic reduction to quadratures \cite{gt05,gt11,ts11,vt12}.

In this note we take a non-Hamiltonian vector from the recent paper by Borisov, Mamaev and Bizyaev \cite{bbm18} and
obtain Poisson bivectors, variables of separation and quadratures using only modern computer software. Our main motivation is to enlarge known collection of deformations of canonical Poisson bivectors appearing both in Hamiltonian and non-Hamiltonian dynamics \cite{bts12,bm14,gt05,gt11}, because  sufficiently large datasets are an integral part of the field of machine learning.

\subsection{  Hamiltonization}

In 1903  Chaplygin found quadratures in the mathematical model of inhomogeneous balanced ball rolling without slipping on a horizontal plane \cite{ch03}. These quadratures for non-Hamiltonian model can be resolved after change of time similar to the well-known Weierstrass change of time in Hamiltonian mechanics \cite{w}. Equations of motion were written in Hamiltonian form only in 2001 \cite{bm01}.

In 1911  Chaplygin introduced  the reducing multiplier method and applied this method to integrate what would later become known as the Chaplygin sleigh \cite{ch11}. He also remarked that his general procedure (using the reducing multiplier) for integrating certain  nonholonomic systems with two degrees of freedom was "interesting from a theoretical standpoint as a direct extension of the Jacobi
method to simple nonholonomic systems."

The first part of Chaplygin's theorem states that in case of nonholonomic systems in two generalized coordinates $(q_1, q_2)$ possessing an invariant measure with density $N(q_1, q_2)$ equations of motion may be written in Hamiltonian form  after the time reparameterization $\mathrm d\tau = N\mathrm dt$. The second part of this theorem says that if a nonholonomic system can be written in Hamiltonian form after  time reparameterization $\mathrm d\tau = f(q_1,q_2)\mathrm dt$, then the original system has an invariant measure. Both functions $N$ and $f$ are  known as the reducing multiplier, or simply the multiplier, see historic remarks and discussion in \cite{bl09,bl11}. The reduced phase space of  Chaplygin's system is isomorphic to the cotangent bundle $T^*Q$
where  reduced equations may be formulated as
  \bq\label{conf-Ham}
\dfrac{\mathrm d z}{\mathrm dt}=N(q_1,q_2)\,\mathcal P_f \,\mathrm d H\,,
 \eq
 where $z=(q_1,q_2,p_1,p_2)$. Roughly speaking Chaplygin considered conformally Hamiltonian vector fields $Z$ associated with Turiel type deformations $\mathcal P_f$ of the canonical Poisson bivector on $T^*Q$ \cite{ts12,tur}. Discussion of symplectic and non-symplectic diffeomorphisms associated with  a conformally Hamiltonian vector field in Hamiltonian mechanics can be found in \cite{mar12}.

After  introduction of Chaplygin's theorem, subsequent research on the theorem  resulted in, among other things, an extension to the quasi coordinate context, a study of the geometry behind the theorem, discoveries of isomorphisms between nonholonomic systems, an example of a system in higher dimensions which was Hamiltonizable through a similar time reparameterization, determination  of necessary conditions for Hamiltonization, study deformations of  Poisson structures in nonholonomic systems, etc. More detailed discussions of various modern methods of the Hamiltonization may be found in \cite{bal12,bbm15,bbm16,bol15,bm14,ko04,fasso18,fed06,bl09,gar18,bl11} and the references therein.

The main advantage of any type of Hamiltonization is that we identify phase space with a Poisson or symplectic manifold. It allows us to study non-Hamiltonian systems using  standard machinery of symplectic geometry.

The main disadvantage of any type of Hamiltonization is that Hamiltonization  only works for a narrow class of nonholonomic systems; even if it works, the reduction to quadratures is not transparent. Another disadvantage is that we do not have an algorithmic procedure for constructing of cotangent bundle $T^*Q$ with generalized coordinates or quasi-coordinates  starting with original physical variables.

\subsection{Separation of variables}

In  \cite{ch11}  Chaplygin discussed a direct extension of the Jacobi method for simple nonholonomic systems. In fact,  we do not need any extensions because the original geometric version of the Jacobi methods is independent of time and, therefore, it is directly  applicable both for Hamiltonian and non-Hamiltonian systems.

In 1837 Jacobi  proved that  $m$ solutions $h_1=H_1(x,y),\ldots,h_m=H_m(x_,y)$ of $m$ separation relations
\bq\label{sep-rel}
\Phi_j(x_j,y_j,h_1,\ldots,h_m)=0\,,\quad j=1,\ldots,m, \quad \mathrm{det}\left[\dfrac{\partial \Phi_j}{\partial h_k}\right]\neq 0\,,
\eq
are in involution with respect to the Poisson bracket
\[
\{H_i,H_j\}_f=0\,,\qquad i,j=1,\ldots,m,
\]
defined by the Poisson bivector
\bq\label{poi-f}
\mathcal P_f=\sum_{j=1}^m f_j(x_j,y_j)\left(
 \dfrac{\partial}{\partial x_j} \wedge \dfrac{\partial}{\partial y_j}
-\dfrac{\partial}{\partial y_j}\wedge \dfrac{\partial }{\partial x_j}\right)\,,
\eq
where $f_j(x_j,y_j)$ are arbitrary functions. Equations  (\ref{sep-rel}) define  curves $X_1,\ldots, X_m$ on a projective plane depending on $m$ parameters $h_1,\ldots,h_m$ so that common level surface of  functions  $H_i(x,y)=h_i$ is a product of these plane curves
\[\mathcal M:\qquad X_1\times X_2\times\cdots X_m\,.\]
 If $\mathcal M$ is a regular Lagrangian submanifold on phase space, we have a completely integrable system, but in generic case $\mathcal M$ is a product of plane curves only.

 Realisations of the common level set  $\mathcal M$ as a product of curves is independent from parameterization of trajectories  living on  $\mathcal M$, i.e. independent of time and of the form of equations of motion. It is pure geometric fact.  We can find this realisation without reduction of equations of motion to the Hamiltonian form.

Compatible Poisson bivectors  $\mathcal P_f$ and $\mathcal P_g$  associated with two sets of functions $f_1,\ldots,f_m$ and $g_1,\ldots,g_m$
are related to each other
\[
\mathcal P_g=N \mathcal P_f
\]
 by the formal  recursion operator
\bq\label{rec-f}
N=\mathcal P_g\mathcal P_f^{-1}=\sum_{j=1}^m
{g_j}{f_j}^{-1}\, L_j\,,
\eq
 where
\[
L_j=
  \dfrac{\partial}{\partial x_j}\otimes d x_j
+\dfrac{\partial}{\partial y_j}\otimes d y_j
\]
 form the so-called algebra of Haantjes  operators with vanishing Nijenhuis torsion.

Functions $H_1,\ldots,H_m$ and compatible Poisson bivectors $\mathcal P_f$ and $\mathcal P_g$  satisfy the equation
\bq\label{contr-f}
\mathcal P_g\mathbf {dH}=\mathcal P_f \mathbf F_{fg}\mathbf {dH}\,, \qquad \mathbf {dH}=(dH_1,dH_2,\cdots,dH_m)\,,
\eq
where $\mathbf F_{fg}$ is the so-called control matrix. Eigenvalues of the control matrix are functions on variables of separation, i.e.
\[
\lambda_j=\lambda_j(x_i,y_i)\,.
\]
If one of the compatible Poisson bivectors $P_f$ or $P_g$ is non-degenerate, we can calculate variables of separation using the recursion operator $N$. If both  bivectors are degenerate, we can calculate variables of separation using the control matrix $\mathbf F$.

Now we are ready to discuss application of the geometric Jacobi method to integration of the equations of motion
\bq\label{gen-eqm}
\dot{z}_i=Z_i(z_1,\ldots,z_n)\,,\qquad i=1,\ldots,n
\eq
which determine a specific mathematical model of some real physical process,  which  means that the number of equations is not very big. First integrals of  vector field $Z$  could be obtained by brute force method, i.e. by solving equation
\bq\label{eq0}
\dot{H}(z)=0\,,
\eq
using some anzats for $H(z)$.  If we find  some first integrals $H_1,\ldots,H_m$, we can try to solve algebraic equations
\bq\label{hp-eq}
\sum {\mathcal P_f}_{ij}\dfrac{\partial H_k}{\partial z_i}\dfrac{\partial H_\ell}{\partial z_j}=0\,,\qquad k,\ell=1,\ldots,m
\eq
with respect to the Poisson bivector  $\mathcal P_f$. Because a'priory these equations  have infinitely many solutions of the form (\ref{poi-f}) we have to restrict the space of solutions, i.e. use a suitable anzats in order to get a partial solution. Instead of (\ref{hp-eq})  we can solve equation
\bq\label{eq1}
Z=\mathrm f_1(z)\mathcal P_f\, \mathrm dH_1+\cdots+\mathrm f_m(z)\mathcal P_f\, \mathrm dH_m\,.
\eq
 It is easy to see that conformally  Hamiltonian vector fields (\ref{conf-Ham}) belong  to a very restricted subspace of  solutions for this equation.

 If we suppose that  equations of motion are reducible to quadratures (completely or partially), then there is also another decomposition
\bq\label{eq2}
Z=\mathrm g_1(z)\mathcal P_g \,\mathrm dH_1+\cdots+\mathrm g_m(z)\mathcal P_g\, \mathrm dH_m\,,
\eq
where $\mathcal P_g$ is a Poisson bivector compatible with  $\mathcal P_f$. A pair of compatible Poisson bivectors determines variables of separation, which allows us to reduce the equations of motion to quadratures. In generic case it could be a complete or partial separation of variables.

Thus, if  equations of motion (\ref{gen-eqm}) can be reduced to quadratures in the framework of the Jacobi method, we have an algorithm of
reduction:
\begin{itemize}
  \item solve equation (\ref{eq0}) with respect to functionally independent first integrals $H_1,\ldots,H_m$;
  \item solve equations (\ref{eq1}-\ref{eq2}) with respect to compatible Poisson bivectors $\mathcal P_f$ and $\mathcal P_g$;
  \item find eigenvalues of the corresponding matrix $F_{fg}$;
  \item calculate quadratures associated with variables of separation.
\end{itemize}
Some results of application of this algorithm in Hamiltonian and non-Hamiltonian mechanics are given in \cite{gt05,gt11,gt16,ts08,ts11,ts11a,vt12}.

The main advantage of the Jacobi method is that using variables of separation, we obtain not only quadratures, but also families  of compatible Poisson brackets,  recursion operators, algebras of Haantjes  operators, master symmetries,  Lax matrices,
new integrable systems and exact discretization of the original equations of motion \cite{ts15,ts17v,ts17c}.

The main disadvantage of the Jacobi method is that we can solve equations (\ref{eq0}) and (\ref{eq1},\ref{eq2})  only using ansatz for a solution. We hope that  selection of the suitable ansatz can be automated using elements of artificial intelligence.

\subsection{Description of the model}
Let as consider  the following equations of motion
\bq\label{eq-m}
\dot{\boldsymbol{\gamma}}=\boldsymbol{\gamma}\times\boldsymbol{\omega}\\,,\qquad
\dot{\mathbf K}=\Omega\boldsymbol{\gamma}\times\mathbf K + (\mathbf K - d\Omega\boldsymbol{\gamma})\times\boldsymbol{\omega}\,.
\eq
Here vectors $\boldsymbol{\gamma}=(\gamma_1,\gamma_2,\gamma_3) $ and $\mathbf K=(K_1,K_2,K_3)$ are variables on the phase space,  $x \times y$  is a vector product in $\mathbb R^3$,   $d$ and $\Omega$ are some  parameters, $ {\mathbf A}=diag(a_1, a_2, a_3)$ is a   diagonal matrix,  and
\[
\boldsymbol{\omega}={\mathbf A}\mathbf K + \frac{(\boldsymbol{\gamma}, {\mathbf A}\mathbf K)}{d^{-1} - (\boldsymbol{\gamma},
{\mathbf A}\boldsymbol{\gamma})}{\mathbf A}\boldsymbol{\gamma}.
\]
These equations describe a partial case of  motion of the inhomogeneous balanced ball rolling without slipping on a horizontal plane rotating with constant angular velocity $\Omega$, see equations (22) in \cite{bbm18}. We use the same notations as in \cite{bbm18},
where the reader can find a complete description of variables  $z=(\g_1,\g_2,\g_3,K_1,K_2,K_3)$, definitions of parameters and a  list of the necessary  references.

According to \cite{bbm18} equations of motion (\ref{eq-m}) possess two geometric  integrals of motion
\bq\label{geom-int}
C_1=\boldsymbol{\gamma}^2=1, \qquad C_2=( \mathbf K,\boldsymbol{\gamma}),
\eq
an integral of motion similar to the Jacobi integral
 \bq\label{jac-int}
H=\frac{1}{2}(\mathbf K, {\mathbf A}\mathbf K) - \frac{\mathbf K^2}{2 d} + \frac{d}{2(1 - d(\boldsymbol{\gamma},
{\mathbf A}\boldsymbol{\gamma}))}({\mathbf A}\mathbf K, \boldsymbol{\gamma})^2\,,
\eq
and an invariant measure
\bq\label{inv-m}
\mu=\rho\, \mathrm d\mathbf K\mathrm d\boldsymbol{\gamma}, \quad\mbox{where}\quad \rho=\big(1 - d(\boldsymbol{\gamma}, {\mathbf A}\boldsymbol{\gamma})\big)^{-\frac{1}{2}}.
\eq

 In Section 2  we present Poisson bivector $\mathcal P$ which allows us to rewrite vector field $Z$ (\ref{eq-m})  in conformally Hamiltonian form
  \[
  Z=\mathrm f(z)\,\mathcal P\mathrm d H(z)
  \]
  at $C_2=0$.  This bivector $\mathcal P$ is  a linear deformation of  the standard Lie-Poisson bivector on  algebra $e^*(3)$    involving non-algebraic functions, i.e. the so-called Turiel type deformation \cite{tur}.

  In Section 3 we discuss a counterpart of the heavy symmetric top at $a_1=a_2$ in (\ref{eq-m}).   In this case we have one more  first integral
  \cite{bbm18}:
\[
H_2=\rho K_3
-\dfrac{da_1\rho\, ( \mathbf K,\boldsymbol{\gamma}) \gamma_3}{da_1-1}+\dfrac{da_1-1}{a_1\sqrt{d(a_1-a_3)}}\,\Omega\ln\left(\sqrt{d (a_1-a_3)}\,\gamma_3+\rho^{-1}\right)\,.
\]
Using this non-algebraic first  integral we can decompose vector field (\ref{eq-m}) into Hamiltonian vector fields
\[
  Z=\mathrm f_1(z)\,\mathcal P'\mathrm d H(z)+\mathrm f_2(z)\,\mathcal P'\mathrm d H_2(z)
  \]
 and find second Poisson bivectors $\mathcal P''$   compatible with $\mathcal P'$ so that
 \[
  Z=\mathrm g_1(z)\,\mathcal P''\mathrm d H(z)+\mathrm g_2(z)\,\mathcal P''\mathrm d H_2(z)
  \]
 It allows us to calculate variables of separation for the equations of motion (\ref{eq-m}) on a computer.

The same  variables of separation may be obtained more easily using  a counterpart of the  Lagrange calculations for the  symmetric heavy top.  In the framework of the Jacobi method these variables of separation determine compatible Poisson brackets,  recursion operators, algebra of Haantjes  operators  and  $2\times2$ Lax matrices for the vector field   (\ref{eq-m}).

\section{Conformally  Hamiltonian vector field at $(\boldsymbol \g,\mathbf K)=0$}
\setcounter{equation}{0}
Let us substitute vector field $Z$ (\ref{eq-m}), its  geometric integrals $C_{1,2}$ (\ref{geom-int})  and  the Jacobi  integral $H$ (\ref{jac-int}) into the following system of algebraic equations
\bq\label{alg-eq}
Z=\mathrm f(z) \mathcal P\mathrm dH\qquad\mbox{and}\qquad \mathcal P\mathrm dC_{1,2}=0\,.
\eq
Desired  Poisson bivector $\mathcal P$ shall also satisfy to the Jacobi identity, i.e.  system of   differential  equations
\bq\label{diff-eq}
[\![\mathcal P,\mathcal P]\!]=0
\eq
coded in a short form using  the Schouten bracket
 \[
[ \![A,B]\!] _{ijk}=-\sum\limits_{m=1}^{dim\, M}\left(B_{mk}\dfrac{\partial A_{ij}}{\partial z_m}
+A_{mk} \dfrac{\partial B_{ij}}{\partial z_m}+\mathrm{cycle}(i,j,k)\right)\,.
\]
In our case $z=(\boldsymbol \g,\mathbf K)$ and  $dim\,  M=6$.

Substituting  linear anzats for  entries of the Poisson bivector
\[
\mathcal P_{ij} =\sum_{m=1}^3 u_{ij}^m(\boldsymbol\g)\,K_m+v_{ij}(\boldsymbol\g)
\]
and function $\mathrm f(z)=\mathrm f(\boldsymbol\g)$ into  (\ref{alg-eq}) one gets an inconsistent system of algebraic equations,  which has solution only at $C_2=(\boldsymbol \g,\mathbf K)=0$. This solution of algebraic equations depends on  arbitrary function $\mathrm f(\boldsymbol \g)$ on variables $\g_1,\g_2$ and $\g_3$.

 Substituting this partial solution into the Jacobi identity and solving the resulting differential equations we obtain the desired Poisson bivector.  It took us only a few seconds to solve both  algebraic and differential equations using one of the modern computer algebra systems.
\begin{prop}
At $\boldsymbol \g^2=1 $ and $(\boldsymbol \g,\mathbf K)=0$ vector field $Z$ (\ref{eq-m}) is a conformally Hamiltonian vector field
\[
Z=\mathrm f(z) \mathcal P \mathrm dH.
\]
with conformal factor depending only on variables $\g_1,\g_2$ and $\g_3$
\[\mathrm f(z)=2\rho\,\delta\,,\]
which is a product of  functions  $\rho$ from  (\ref{inv-m}) and
\[
\delta= (1-d a_1)a_2 a_3\g_1^2+ (1-d a_2)a_1 a_3\g_2^2+ (1-d a_3)a_1 a_2\g_3^2\,.
\]
 \end{prop}
The proof consists of straightforward verification of algebraic and differential equations  (\ref{alg-eq}-\ref{diff-eq}) by using an explicit form of  Poisson bivector $\mathcal P$.

 In variables $z=(\boldsymbol\g, \mathbf K)$ bivector ${\mathcal P}$  is equal to
\bq\label{poi-K}
\mathcal P=\frac{d}{2\delta\mu}
\left(
  \begin{array}{cc}
    0 &\boldsymbol\Upsilon \\ \\
    - \boldsymbol\Upsilon^\top &\Omega\, \boldsymbol{\Gamma}\\
  \end{array}
\right)
+\frac{d\mu}{2\delta}\left(
  \begin{array}{cc}
    0 & 0\\ \\
   0 & \boldsymbol{\Pi}\\
  \end{array}
\right)
\,,
\eq
where $\boldsymbol\Upsilon$  is the following $3\times 3$ matrix
 \[\boldsymbol\Upsilon=\boldsymbol{\Gamma}_3-\boldsymbol{\Gamma}{\mathbf A} (d{\mathbf A}-{\mathbf Id} )^{-1}
\,,\]
matrix $\boldsymbol \Gamma$ is equal to
\[
\mathbf \Gamma=\left(
                 \begin{array}{ccc}
                   0 & \g_3 & -\g_2 \\
                   -\g_3 & 0 & \g_1 \\
                   \g_2 & -\g_1 & 0
                 \end{array}
               \right)
\]
and
\[
 \boldsymbol{\Gamma}_3=\left(
                         \begin{array}{ccc}
                          \frac{ {\g }_1 {\g }_2 {\g }_3 (a_3-a_2)}{(d a_3-1) (d a_2-1)} &
                          \frac{ {\g }_2^2{\g }_3 (a_3-a_2)}{(d a_3-1) (d a_2-1)} &
                          \frac{{\g }_2 {\g }_3^2 (a_3-a_2)}{(d a_3-1) (d a_2-1)} \\ \\
                          \frac{{\g }_1^2 {\g }_3(a_1-a_3)}{(d a_3-1)(d a_1-1)} &
                          \frac{{\g }_1 {\g }_2 {\g }_3 (a_1-a_3)}{(d a_3-1) (d a_1-1)} &
                         \frac{{\g }_1 {\g }_3^2(a_1-a_3)}{(d a_3-1) (d a_1-1)}  \\ \\
                        \frac{{\g }_1^2 {\g }_2 (a_2-a_1)}{(d a_2-1)(d a_1-1)}   &
                         \frac{{\g }_1 {\g }_2^2(a_2-a_1)}{(d a_2-1)(d a_1-1)}  &
                         \frac{{\g }_1 {\g }_2 {\g }_3(a_2-a_1)}{(d a_2-1)(d a_1-1)}  \\
                         \end{array}
                       \right)\,,
 \]
so that
\[
 \boldsymbol{\Gamma}_3 \boldsymbol{\Gamma}=0\,.
\]
Skew-symmetric $3\times 3$ matrix $\boldsymbol\Pi$ has a more cumbersome form
\[
\boldsymbol{\Pi}=\frac{1}{\det (d{\bf A}-{\bf Id}) }
\left(
  \begin{array}{ccc}
    0 &
    \frac{\alpha_1 d {\g }_1 {\g }_2 {\g }_3({\boldsymbol\g }\times {\bf K})_3+\beta_3{\bf K}_3}{{\g }_1^2+{\g }_2^2} &
    \frac{\alpha_2 d {\g }_1 {\g }_2 {\g }_3 ({\boldsymbol\g }\times {\bf K})_2+\beta_2{\bf K}_2}{{\g }_1^2+{\g }_3^2}\\ \\
   *  & 0 &
     \frac{\alpha_1 d {\g }_1 {\g }_2 {\g }_3 ({\boldsymbol\g }\times {\bf K})_1+\beta_1{\bf K}_1}{{\g }_2^2+{\g }_3^2} \\ \\
     *& * & 0 \\
  \end{array}
\right)\,,
\]
where
\[\begin{array}{rcl}
\alpha_3&=& (a_2-a_1) (d (a_2-a_3) (a_3-a_1) {\g }_3^2+a_3 (1-d a_2) (1-d a_1))\,,\\
\\
\alpha_2&=& (a_1-a_3) (d (a_2-a_3) (a_1-a_2) {\g }_2^2+a_2 (1-d a_3) (1-d a_1))\,,\\
\\
\alpha_1&=& (a_2-a_3) (d (a_1-a_3) (a_2-a_1) {\g }_1^2+a_1(1-d a_3)  (1-d a_2))
\end{array}
\]
and
\[\begin{array}{rcl}
\beta_1&=&\scriptstyle d (a_1 -a_2 )^2 (1-d a_3)\g_2^6+d (a_1-a_3 )^2 (1-d a_2 )\g_3^6\\
&-&\scriptstyle
d (a_1 -a_2 ) (d a_1  a_2 +2 d a_1  a_3 -2 d a_2  a_3 -d a_3 ^2-3 a_1 +a_2 +2 a_3 )\g_2^4\g_3^2\\
&-&\scriptstyle
d (a_1-a_3) (2 d a_1  a_2 +d a_1  a_3 -d a_2 ^2-2 d a_2  a_3 -3 a_1 +2 a_2 +a_3 )\g_2^2\g_3^4\\
&+&\scriptstyle
(a_1 -a_2) (2 d^3 a_1  a_2  a_3 -3 d^2 a_2  a_3 -2 d a_1 +d a_2 +d a_3 +1)\g_2^4\\
&+&\scriptstyle
(a_1-a_3) (2 d^3 a_1  a_2  a_3 -3 d^2 a_2  a_3 -2 d a_1 +d a_2 +d a_3 +1)\g_3^4\\
&+&\scriptstyle
(2 a_1 -a_2 -a_3 ) (2 d^3 a_1  a_2  a_3 -3 d^2 a_2  a_3 -2 d a_1 +d a_2 +d a_3 +1)\g_2^2\g_3^2\\
&+&\scriptstyle
(1-d a_1) (2 d^2 a_1  a_2  a_3 -d^2 a_2 ^2 a_3 -d a_1  a_3 -d a_2  a_3 -a_1 +a_2 +a_3 )\g_2^2\\
&+&\scriptstyle
(1-d a_1) (2 d^2 a_1  a_2  a_3 -d^2 a_2  a_3 ^2-d a_1  a_2 -d a_2  a_3 -a_1 +a_2 +a_3 )\g_3^2\,,
\end{array}
\]
\[\begin{array}{rcl}
\beta_2&=&\scriptstyle
d (a_1 -a_2 )^2 (d a_3 -1)\g_1^6+d (a_2 -a_3 )^2 (d a_1 -1)\g_3^6\\
&-&\scriptstyle
d (a_1 -a_2 ) (d a_1  a_2 -2 d a_1  a_3 +2 d a_2  a_3 -d a_3 ^2+a_1 -3 a_2 +2 a_3 )\g_1^4\g_3^2\\
&-&\scriptstyle
d (a_2 -a_3 ) (d a_1 ^2-2 d a_1  a_2 +2 d a_1  a_3 -d a_2  a_3 -2 a_1 +3 a_2 -a_3 )\g_1^2\g_3^4\\
&+&\scriptstyle
(a_1 -a_2 ) (2 d^3 a_1  a_2  a_3 -3 d^2 a_1  a_3 +d a_1 -2 d a_2 +d a_3 +1)\g_1^4\\
&+&\scriptstyle
(a_3 -a_2 ) (2 d^3 a_1  a_2  a_3 -3 d^2 a_1  a_3 +d a_1 -2 d a_2 +d a_3 +1)\g_3^4\\
&+&\scriptstyle
(a_1 -2 a_2 +a_3 ) (2 d^3 a_1  a_2  a_3 -3 d^2 a_1  a_3 +d a_1 -2 d a_2 +d a_3 +1)\g_1^2\g_3^2\\
&+&\scriptstyle
(d a_2 -1) (2 d^2 a_1  a_2  a_3-d^2 a_1 ^2 a_3 -d a_1  a_3 -d a_2  a_3+-a_1 -a_2+a_3 )\g_1^2\\
&+&\scriptstyle
(d a_2 -1) (2 d^2 a_1  a_2  a_3 -d^2 a_1  a_3 ^2-d a_1  a_2 -d a_1  a_3 +a_1 -a_2 +a_3 )\g_3^2\,,
\end{array}
\]
\[\begin{array}{rcl}
\beta_3&=&\scriptstyle
d (a_1-a_3 )^2 (d a_2 -1)\g_1^6+d (a_2 -a_3 )^2 (d a_1 -1)\g_2^6\\
&-&\scriptstyle
d (a_3 -a_1 ) (2 d a_1  a_2 -d a_1  a_3 +d a_2 ^2-2 d a_2  a_3 -a_1 -2 a_2 +3 a_3 )\g_1^4\g_2^2\\
&-&\scriptstyle
d (a_3-a_2 ) (d a_1 ^2+2 d a_1  a_2 -2 d a_1  a_3 -d a_2  a_3 -2 a_1 -a_2 +3 a_3 )\g_1^2\g_2^4\\
&+&\scriptstyle
(a_1-a_3 ) (2 d^3 a_1  a_2  a_3 -3 d^2 a_1  a_2 +d a_1 +d a_2 -2 d a_3 +1)\g_1^4\\
&+&\scriptstyle
(a_2 -a_3 ) (2 d^3 a_1  a_2  a_3 -3 d^2 a_1  a_2 +d a_1 +d a_2 -2 d a_3 +1)\g_2^4\\
&+&\scriptstyle
(a_1 +a_2 -2 a_3 ) (2 d^3 a_1  a_2  a_3 -3 d^2 a_1  a_2 +d a_1 +d a_2 -2 d a_3 +1)\g_1^2\g_2^2\\
&+&\scriptstyle
(1-d a_3 ) (d^2 a_1 ^2 a_2 -2 d^2 a_1  a_2  a_3 +d a_1  a_2 +d a_2  a_3 -a_1 -a_2 +a_3 )\g_1^2\\
&+&\scriptstyle
(1-d a_3 ) (d^2 a_1  a_2 ^2-2 d^2 a_1  a_2  a_3 +d a_1  a_2 +d a_1  a_3 -a_1 -a_2 +a_3 )\g_2^2\,.
\end{array}
\]
It's easier to get this solution on a computer than to write it out.

At $\boldsymbol\g^2=1$ and $(\boldsymbol\g,\mathbf K)=0$ we can simplify these expressions by using algebraic transformation  variables $ K_i\to  L_i$ defined by equations of the form
\bq\label{var-L}
\begin{array}{rcl}
{ L}_1&=&\frac{2\rho(1-da_1)
\Bigl(\bigl(a_2 a_3 ({\g }_2^2+{\g }_3^2) d-a_3{\g }_2^2 -a_2{\g }_3^2 \bigr) { K}_1-(a_3-a_2) {\g }_1 {\g }_2 {\g }_3 \bigl( {\boldsymbol\g }\times {\mathbf K}\bigr)_1\Bigr)
}{d({\g }_2^2+{\g }_3^2)}\,,\\
\\
{L}_2&=&\frac{2\rho(1-da_2)
\Bigl(\bigl(a_1 a_3 ({\g }_1^2+{\g }_3^2) d-a_3 {\g }_1^2-a_1 {\g }_3^2\bigr) { K}_2-(-a_3+a_1) {\g }_1 {\g }_2 {\g }_3 \bigl( {\boldsymbol\g }\times {\mathbf K}\bigr)_2\Bigr)
}{d({\g }_1^2+{\g }_3^2)}\,,
\\
\\
{ L}_3&=&\frac{2\rho(1-da_3)
\Bigl(\bigl(d a_1 a_2 ({\g }_1^2+{\g }_2^2)-{\g }_1^2 a_2-a_1 {\g }_2^2\bigr) { K}_3-(a_2-a_1) {\g }_1 {\g }_2 {\g }_3\bigl( {\boldsymbol\g }\times {\mathbf K}\bigr)_3
\Bigr)}{d({\g }_1^2+{\g }_2^2)}\,.
\end{array}
\eq
Here $\rho$ (\ref{inv-m}) is an algebraic function of variables $\g_1,\g_2$ and $\g_3$.

In variables $z=(\boldsymbol \g,\mathbf L)$  Poisson bivector $\mathcal P$ (\ref{poi-K}) becomes  a quite visible object
\begin{eqnarray}\label{poi-L}
\mathcal P&=&\left( \begin{array}{rrrrrr}
          0 & 0 & 0 & 0 & {\g }_3 & -{\g }_2 \\
          0 & 0 & 0 & -{\g }_3 & 0 & {\g }_1 \\
          0 & 0 & 0 & {\g }_2 & -{\g }_1 & 0 \\
          0 & {\g }_3 & -{\g }_2 & 0 & { L}_3 & -{ L}_2 \\
          -{\g }_3 & 0 & {\g }_1 & -{ L}_3 & 0 & { L}_1 \\
          {\g }_2 & -{\g }_1 & 0 & {L}_2 & -{L}_1 & 0 \\
        \end{array}
      \right) \\
   \nonumber   \\
&+&\frac{2\rho (da_1-1)(da_2-1)(da_3-1)\Omega}{d}
\left(
        \begin{array}{rrrrrr}
          0 & 0 & 0 & 0 & 0 & 0 \\
          0 & 0 & 0 & 0 & 0 & 0 \\
          0 & 0 & 0 &0 & 0  & 0 \\
          0 & 0 & 0 & 0 & {\boldsymbol\gamma}_3 & -{\boldsymbol\gamma}_2 \\
          0& 0 & {\g }_1 & -{\boldsymbol\gamma}_3 & 0 & {\boldsymbol\gamma}_1 \\
          0 &0 & 0 & {\boldsymbol\gamma}_2 & -{\boldsymbol\gamma}_1 & 0 \\
        \end{array}
      \right)\,.\nonumber
    \end{eqnarray}
Below we prove that this bivector is  a non algebraic deformation of the Lie-Poisson bivector on $e^*(3)$
 \bq\label{poi-can}
\mathcal P_0=\left(
                       \begin{array}{cc}
                         0 & \mathbf\Gamma \\
                       -\mathbf\Gamma^\top & \mathbf M \\
                       \end{array}
                     \right)
\,,\eq
where
\[
\mathbf \Gamma=\left(
                 \begin{array}{ccc}
                   0 & \g_3 & -\g_2 \\
                   -\g_3 & 0 & \g_1 \\
                   \g_2 & -\g_1 & 0
                 \end{array}
               \right)\,,\qquad
\mathbf M=\left(
                 \begin{array}{ccc}
                   0 & M_3 & -M_2 \\
                   -M_3 & 0 & M_1 \\
                   M_2 & -M_1 & 0
                 \end{array}
               \right)\,,
\]
 It means that we cannot reduce bivector  $\mathcal P$ (\ref{poi-L}) to the  Lie-Poisson  bivector $\mathcal P_0$ (\ref{poi-can})  using only algebraic transformations of variables.

\subsection{Deformation of the canonical Poisson bivector linear in momenta}

Let $Q$ be an $n$-dimensional smooth manifold endowed with (1,1) tensor field $\Lambda (q)$ with vanishing Nijenhuis torsion and with vector field $\nu(q)$. Canonical  Poisson bracket on  its cotangent bundle $T^*Q$
\[
\{q_i,q_j\}=0\,,\qquad \{q_i,p_j\}=\delta_{ij}\,,\qquad \{p_i,p_j\}=0\,,
\]
where $q_i,p_i$ are fibered coordinates, after linear transformation  of momenta
\[
p_j\to \sum_{i=1}^n \Lambda^i_j(q)p_i+\nu_j(q)
\]
looks like
\[
\{q_i,q_j\}'=0\,,\qquad \{q_i,p_j\}'=\Lambda^i_j\,,\qquad
\{p_i,p_j\}'=\sum_{k=1}^n \left(
\dfrac{\partial \Lambda^k_i}{\partial q_j}-\dfrac{\partial \Lambda^k_j}{\partial q_i}
\right)p_k+\dfrac{\partial \nu_j}{\partial q_i}-\dfrac{\partial \nu_i}{\partial q_j}\,.
\]
It is the so-called Turiel type  deformation of the canonical Poisson bracket on $T^*Q$, see \cite{tur,ts12} and references within.

Bivector $\mathcal P$ (\ref{poi-L}) determines standard Poisson brackets $\{\g_i,\g_j\}=0$ and $\{\g_i,L_j\}=\varepsilon _{ijk}\g_k$ and non-standard brackets $\{L_i,L_j\}$ between momenta.  Let us consider a change of variables
\bq\label{var-M}
L_i\to M_i=M_i+\eta_i(\boldsymbol\gamma)\,,
\eq
which reduces these non-standard Poisson brackets to canonical form
\bq\label{eq-LM}
\{M_1,M_2\}=M_3\,,\qquad \{M_2,M_3\}=M_1\,,\qquad \{M_3,M_1\}=M_2
\eq
at
\[
(\boldsymbol \g, \mathbf M)=\eta_1\g_1+\eta_2\g_2+\eta_3\g_3=0\qquad\mbox{and}\qquad
\g_1^2+\g_2^2+\g_3^2=1.
\]
These algebraic equations hold if we use  the following anzats
\[\eta_1=\g_2\eta_4-\dfrac{\g_1}{\g_1^2+\g_2^2}\,\g_3\eta_3\,,\qquad
\eta_2=-\g_1\eta_4-\dfrac{\g_2}{\g_1^2+\g_2^2}\,\g_3\eta_3\,,
\]
for  functions $\eta_{1,2}$. We substitute this anzats into (\ref{eq-LM}) and solve the resulting differential equations on a computer.

First partial solution of the equations (\ref{eq-LM}) reads as $\eta_3=0$ and
\[\begin{array}{c}
\eta_4=\frac{2(1-d a_1)(1-d a_2)(1-d a_3)\Omega}{d^{3/2}\sqrt{a_1-a_2}\sqrt{\g_3^2-1}}\,
F\left(
\sqrt{
\frac{d(a_2-a_1)\g_2^2}{d(a_1-a_2)\g_3^2-da_1+1}
},
\sqrt{
\frac{d(a_1-a_2)\g_3^2-da_1+1}{d(a_2-a_2)(\g_1^2+\g_2^2)}
}
\right)\,,
\end{array}
\]
where $F$ is an incomplete elliptic integral of the first kind.

Second  partial solution of the equations (\ref{eq-LM}) looks like $\eta_4=0$ and
\[\eta_3=\dfrac{2(1-d a_1)(1-d a_2)(1-d a_3)\Omega}{
d^{3/2}\kappa}\,\ln\left(\sqrt{d}\,\kappa{\boldsymbol \gamma}_3+\dfrac{1}{\rho}\right)\,,
\]
where function  $\rho$ defines the invariant measure (\ref{inv-m}) and
\[
\kappa=\sqrt{\dfrac{(a_1-a_3){\boldsymbol \gamma}_1^2+(a_2-a_3){\boldsymbol \gamma}_2^2}{{\boldsymbol \gamma}_1^2+{\boldsymbol \gamma}_2^2}}\,.
\]
Thus, in variables $z=(\boldsymbol \g,\mathbf M)$  vector field $Z$ (\ref{eq-m}) becomes  a conformally Hamiltonian vector field
\[
Z=\mathrm f(\boldsymbol \g) \mathcal P\mathrm dH
\]
 defined by  Hamiltonian $H$ (\ref{jac-int}) which is a non-algebraic function in  variables $\mathbf M$ (\ref{var-M}) on the  cotangent bundle of  two-dimensional sphere $T^*\mathbb S^2$

This result was obtained by directly solving algebraic and differential equations on a computer.  Now we can apply this result in order to prove that bivector  $\mathcal P$ is the Turiel type deformation of canonical Poisson brackets. As we identify phase space with $T^*\mathbb S^2$ we can introduce spherical coordinates and momenta
\[
\phi =\arctan\g_1/\g_2\,,\quad  p'_\phi = -K_3,\quad
\theta =\arccos \g_3,\quad  p'_\theta = -\dfrac{\g_2 K_1-\g_1 K_2}{\sqrt{\g_1^2+\g_2^2}}\,.
\]
\begin{prop}
Transformation of momenta
\[\begin{array}{rcl}
p_\phi&=&\Lambda^1_1(\phi,\theta)p'_\phi+\Lambda^2_1(\phi,\theta)p'_\theta+\nu_1(\phi,\theta)\,,\\
\\
p_\theta&=&\Lambda^1_2(\phi,\theta)p'_\phi+\Lambda^2_2(\phi,\theta)p'_\theta+\nu_2(\phi,\theta)
\end{array}
\]
reduces the Poisson bracket associated with bivector $\mathcal P$ (\ref{poi-K}) to a canonical Poisson bracket on the cotangent bundle of two-dimensional sphere
\[
\{\phi,\theta\}=0\,,\qquad\{ \phi,p_\phi \}=\{\theta,p_\theta \}=1\,,\qquad
\{ p_\phi,p_\theta \}=0\,,
\]
if
\[
\Lambda=\frac{2 (da_3-1)\rho}{d}
\left(
  \begin{array}{cc}
  (a_1-a_2)\cos^2\phi-da_1a_2+a_2 &\frac{ (a_1-a_2)\sin2\phi\sin2\theta}{4} \\ \\
    \frac{(a_1-a_2)\sin2\phi\cos\theta}{2\sin\theta} & \frac{\lambda}{da_3-1}
    \end{array}
\right)
\]
where
\[
\lambda=
-(da_3-1)(a_1-a_2)\cos^2\theta\cos^2\phi+
(da_2-1)(a_1-a_3)\cos^2\theta-(da_1-1)(da_2-1)a_3
\]
and functions $\nu_{1,2}(\phi,\theta)$ satisfy differential equation
\[
\dfrac{\partial \nu_1}{\partial \theta}-\dfrac{\partial \nu_2}{\partial \phi}
 =2\Omega(1-d a_1)(1-d a_2)(1-d a_3)\rho\sin\theta\,.
\]
Partial solutions of this equation are incomplete elliptic integral of the first kind and  logarithmic function, which have been obtained  by brute force method before.
\end{prop}
 It is easyto prove that (1,1) tensor field $\Lambda (q)$ has zero Nijenhuis torsion and, therefore,
  bivector $\mathcal P$ (\ref{poi-K}) is  a Turiel type deformation of the canonical Poisson bivector.

In nonholonomic mechanics functions $L^i_j(q)$ and $\nu_j(q)$ are usually algebraic functions on configuration space $Q$. For instance, see linear deformations appearing for:
  \begin{itemize}
    \item reduced  motion of the Chaplygin ball on the plane \cite{ts11a};
      \item  reduced  motion of the Chaplygin ball on the sphere  \cite{ts12rd};
    \item nonholonomic Veselova system and its generalizations \cite{ts12b};
    \item reduced motion of the Routh ball on the plane \cite{bts12};
    \item nonholonomic motion of a body of revolution on the plane \cite{ts14b};
    \item nonholonomic motion of a homogeneous ball on the surface of revolution \cite{ts14b};
    \item nonholonomic  Heisenberg type systems \cite{gt16};
    \item other non-Hamiltonian systems associated with Turiel type deformations \cite{ts12}.
  \end{itemize}
In contrast with  these systems with two degrees of freedom  one
gets non-algebraic deformations involving elliptic integrals or logarithms for  reduced  motion of the Chaplygin ball on the rotating plane (\ref{eq-m}).

 In fact, it is the main result because it allows us to essentially enlarge the list of possible Turiel type deformations  appearing in mathematical description of real physical systems. We suppose that similar non-algebraic deformations of the canonical Poisson brackets appear also in other models of rigid body  motion on rotating surfaces \cite{e44,ll86,rou55,tz25,z17}.

Non-algebraic  deformations for the non-Hamiltonian systems with three degrees of freedom are also discussed in \cite{bol15,ts14a}.

\section{Sum of the Hamiltonian vector fields }
\setcounter{equation}{0}

 At $a_1=a_2$ vector field $Z$ (\ref{eq-m}) has a formal integral of motion
\bq\label{ham-h2}
H_2=\rho K_3
--\dfrac{da_1\rho\, (\boldsymbol{\gamma}, \mathbf K) \gamma_3}{da_1-1}+\dfrac{da_1-1}{a_1\sqrt{d(a_1-a_3)}}\,\Omega\ln\left(\sqrt{d (a_1-a_3)}\,\gamma_3+\rho^{-1}\right)\,,
\eq
with a logarithmic term, see discussion in   \cite{bbm18}.

\begin{prop}
At $a_1=a_2$ vector field $Z$ (\ref{eq-m}) is a sum of two Hamiltonian vector fields
\bq\label{dec-eq1}
Z={\mathrm f}_1(z)\mathcal P'\,\mathrm  dH+{\mathrm f}_2(z)\mathcal P'\, \mathrm dH_2
\eq
with coefficients
\[ \mathrm f_1=2\rho\,\delta\,,\qquad\mbox{and}\qquad \mathrm f_2=\dfrac{2a_1(a_1-a_3)\g_3\cdot (\boldsymbol \g,\mathbf K)}{d}\,.
\]
Here  $H$ is a Jacobi integral and Poisson bivector $P'$ is equal to
\bq\label{ch-poi2}
\mathcal P'=\mathcal P+\dfrac{ d^2\rho\cdot(\boldsymbol \g,\mathbf K)}{2\Bigl((a_3-a_1)\g_3^2+(da_1-1)a_3\Bigr)(da_1-1)}
\begin{pmatrix}
                          0 &0 & 0 & 0 & 0 & 0 \\
                          0 & 0 & 0 & 0 & 0 & 0 \\
                          0 & 0 & 0 & 0 & 0 & 0 \\
                          0 & 0 & 0 & 0 & \sigma_3 & -\sigma_2 \\
                          0 &0 & 0 & -\sigma_3 & 0 & \sigma_1 \\
                          0 & 0 & 0 & \sigma_2 & -\sigma_1 & 0 \\
                        \end{pmatrix}\,,
\eq
where
\[
\sigma_1=\left(a_1-\dfrac{(a_1-a_3)\g_3^2}{1-\g_1^2}\right)\g_1\,,\qquad
\sigma_2=\left(a_1-\dfrac{(a_1-a_3)\g_3^2}{1-\g_2^2}\right)\g_2\,,
\]
and
\[
\sigma_3=\left(a_1a_3(da_1-1) -(a_1-a_3)^2\g_3^2+\dfrac{a_3-a_1}{d}\right)\dfrac{\g_3}{da_3-1}\,.
\]
\end{prop}
The proof is a straightforward calculation.

In this case we cannot use any type of Hamiltonization for  reduction of equations of motion to a Hamiltonian form.
Nevertheless, we can calculate variables of separation for equations of motion (\ref{eq-m})
directly solving equation  (\ref{eq2}) in the framework of the Jacobi method.

Because we know the solution (\ref{dec-eq1}) of equation (\ref{eq1}) we have to solve  (\ref{eq2})  together with equation
\[
[\![\mathcal P',\mathcal P'']\!]=0\,,
\]
which guarantees compatibility of bivectors $ \mathcal P'$ and $\mathcal P''$. Solving these equations on a computer we
use the following anzats:
 \begin{itemize}
   \item entries of $\mathcal P''$ are linear functions on momenta;
   \item  first coefficient $\mathrm g_1$ is a function on $\g_3$ similar to (\ref{dec-eq1} );
   \item second coefficient $\mathrm g_2$ is a linear function in momenta similar to (\ref{dec-eq1} ).
 \end{itemize}
 After a few seconds one gets the following answer
\[
\mathcal P''=
\left(
  \begin{array}{cc}
    0 &\boldsymbol\Upsilon'' \\ \\
    - {\boldsymbol\Upsilon''}^\top & \boldsymbol{\Pi}''\\
  \end{array}
\right)\,,
\]
where
\[
\boldsymbol\Upsilon''=
\left(
  \begin{array}{ccc}
     -\frac{\g_1 \g_2\g_3}{\g_1^2+\g_2^2}\left(\frac{1}{\rho}-\frac{1}{a_1(\g_1^2+\g_2^2)}\right) &
   -\frac{\g_3}{\g_1^2+\g_2^2}\left(\frac{\g_2^2}{\rho}+\frac{\g_1^2}{a_1(\g_1^2+\g_2^2)}\right)&
    \frac{\g_2}{\rho} \\ \\
    \frac{\g_3}{\g_1^2+\g_2^2}\left(\frac{\g_1^2}{\rho}+\frac{\g_2^2}{a_1(\g_1^2+\g_2^2)}\right) &
      \frac{\g_1 \g_2\g_3}{\g_1^2+\g_2^2}\left(\frac{1}{\rho}-\frac{1}{a_1(\g_1^2+\g_2^2)}\right) &
      -\frac{\g_1}{\rho}\\ \\
       -\frac{\g_2}{a_1(\g_1^2+\g_2^2)}&\frac{ \g_1}{a_1(\g_1^2+\g_2^2)} &0  \\
     \end{array}
\right)
\]
and entries of the skew symmetric matrix $\boldsymbol{\Pi}''$ are equal to
\[\begin{array}{rcl}
\boldsymbol{\Pi}''_{12}
&=&\frac{\g_3}{\g_1^2+\g_2^2} \left(-\frac{(\g_1 K_1+\g_2 K_2)}{\rho}
+\frac{(da_1-1) \Omega-a_1 K_3}{a_1^2}
+\frac{\rho^2\left(da_3 \g_3 K_3-\frac{(x_1K_1+x_2K_2) (da_3 \g_3^2-1)}{\g_1^2+\g_2^2}\right)}{a_1}
\right)\,,\\
\\
\boldsymbol{\Pi}''_{13}
&=&\phantom{-}\frac{K_2}{\rho} -\left(\frac{(d a_1-1) \Omega}{a_1}
-{d \Bigl(a_1( \g_1K_1+\g_2K_2)+a_3\g_3K_3\Bigr)\rho^2}\right)\frac{\g_2}{a_1(\g_1^2+\g_2^2)}\,,
\\ \\
\boldsymbol{\Pi}''_{23}
&=&-\frac{K_1}{\rho}+\left(\frac{ (da_1-1) \Omega}{a_1} +d  \Bigl(a_1(\g_1 K_1+x_2K_2)+a_3\g_3K_3\Bigr) \rho^2\right)\frac{\g_1}{a_1(\g_1^2+\g_2^2)}\,.
\end{array}
\]
In this case coefficients in decomposition  (\ref{eq2}) are equal to
\[
\mathrm g_1(z)=\dfrac{a_1d(1-\g_3^2)}{2(1-da_1)}
\]
and
\[\begin{array}{rcl}
\mathrm g_2(z)&=&\left(\dfrac{a_1^2 (\g_1^2+\g_2^2)}{(d a_1-1) \rho}+da_1^2 a_3  (\g_1^2+\g_2^2) \rho-a_3 (d a_1-1) \rho^2\right)K_3\\
\\
&-&a_1 \g_3\left(\dfrac{a_1}{(d a_1-1) \rho}+\dfrac{da_1^2  (d a_3-1)\rho}{(d a_1-1)}-\dfrac{(d a_3-1) \rho^2}{\g_1^2+\g_2^2}\right) (\g_1 K_1+\g_2 K_2)\,.
\end{array}
\]
Both bivectors $\mathcal P'$ and $\mathcal P''$ are degenerate and, therefore, we cannot determine recursion operator (\ref{rec-f}), but we can easily calculate control matrix $\mathbf F$ (\ref{contr-f}) and its eigenvalues, which are desired variables of separation. In our case we have to solve algebraic equations
\[\begin{array}{rcl}
\mathcal P'\mathrm dH&=&\mathbf F_{11}\mathcal P''\mathrm dH+\mathbf F_{12}\mathcal P''\mathrm dH_2\,,\\
\\
\mathcal P'\mathrm dH_2&=&\mathbf F_{21}\mathcal P''\mathrm dH+\mathbf F_{22}\mathcal P''\mathrm dH_2
\end{array}
\]
with respect to four entries of matrix $\mathbf F$. In our case $\mathbf F_{21}=0$
and we can easily calculate eigenvalues of $\mathbf F$
\[
\lambda_1=\mathbf F_{21}==\dfrac{d}{2a_1(da_3-1)(da_1-1)}\qquad\mbox{and}\qquad
\lambda_2=\dfrac{ d(1-\g_3^2) a_1\sqrt{d(a_1-a_3)\g_3^2-a_1d+1}}
{2(da_1-1)(da_1a_3-(a_1-a_3)\g_3^2-a_3)}\,.
\]
There is only one nontrivial eigenvalue $\lambda_2$  which is the desired variable of separation. Now we have to validate that
our algorithm yields quadrature
\[
\dfrac{\mathrm d\lambda_2}{\mathrm dt}=F(\lambda_2)\,,
\]
 i.e. an equation which includes only  variable $\lambda_2$ and  its differential.  This quadrature will be discussed in the following subsection.

Of course, this  variable of separation may be obtained by directly using  symmetries of the ball similar to  the Lagrange approach to symmetric heavy top. More complicated, but algorithmic computer calculations of variables of separation were given in order to underline the importance of studying  equations (\ref{eq1}-\ref{eq2})  in nonholonomic mechanics.

\subsection{Separation of variables}
It is easy to  prove that  equation of motion
\bq\label{ch-dot-x3}
\dot{\g}_3^2=a_1^2\left(\g_2M_1-\g_1M_2\right)^2=F(\g_3)
\eq
on  common level surface of the first integrals
\bq\label{ch-comm}
\qquad C_1=1\,,\qquad C_2=\ell\,,\qquad H_2=k\,,\qquad 2H=h
\eq
includes only one variable with its own differential. Here
\bq\label{ch-fx}
\begin{array}{rcl}
F(x)&=&
\dfrac{a_1^2 d (x^2-1)}{d a_1-1}\,h
+\dfrac{a_1^2(x^2-1) (dx^2(a_1-a_3)+2 d^2 a_1 a_3-2 d a_1-d a_3+1)}{(d a_3-1) (d a_1-1)^2}\, \ell^2\\
\\
&-&\left(
\dfrac{a_1x\sqrt{dx^2(a_1-a_3)-da_1+1}}{(da_1-1)\sqrt{da_3-1}}\,\ell-a_1\sqrt{da_3-1}\,k\right.\\
\\
&+&\left.
\dfrac{(da_1-1)\sqrt{da_3 -1}\,\ln
\Bigl(\sqrt{d(a_1-a_3)}x+\sqrt{dx^2(a_1-a_3)-da_1+1}\Bigr) \Omega}{\sqrt{d(a_1-a_3)}}
\right)^2\,.
\end{array}
\eq
Thus,  at $a_1=a_2$ solution of the six equations of motion (\ref{eq-m}) is reduced to one nontrivial quadrature
\[
\int^{\g_3} \dfrac{\mathrm d x}{\sqrt{F(x)}}=t
\]
involving non-algebraic function $F(x)$ (\ref{ch-fx}).   We hope that study of such non-Abelian quadratures can be carried out by means other than numerical simulations.

Following Lagrange and Jacobi we can introduce  two pairs of independent variables of separation
\bq\label{ch-div}
x_1=\g_3,,\quad y_1=a_1(\g_1K_2-\g_2K_1)\,,\qquad\mbox{and}
\qquad
x_2=\phi(\g_1/\g_2)\,,\quad y_2=H_2\,.
\eq
so that
\[
\{x_1,x_2\}'=\{y_1,y_2\}'=\{y_1,x_2\}'=\{x_1,y_2\}'=0
\]
and
\bq\label{ch-f12}
\begin{array}{rcl}
\{x_1,y_1\}'&=&f_1(x_1)\equiv\dfrac{a_1 \sqrt{d (a_1-a_3) x_1^2-a_1 d+1} (x_1^2-1)}{d (a_1-a_3) \Bigl(a_3(d a_1-1)-(a_1-a_3)x_1^2\Bigr)} \,,\\
\\
\{x_2,y_2\}'&=& f_2(x_2)\equiv\dfrac{\psi(x_2)}{da_1(a_1-a_3)(da_3-1)}\,,\qquad \psi=\dfrac{\mathrm d\,\phi(\g_1/\g_2)}{\mathrm d (\g_1/\g_2)}\,.\\
\end{array}
\eq
Here $\phi(\g_1/\g_2)$ is an arbitrary function and $\{.,.\}'$ is the Poisson bracket associated with the Poisson  bivector $\mathcal P'$ (\ref{ch-poi2}).

In the framework of the Jacobi method we  identify these variables of separation with affine coordinates of divisors.
First divisor $P_1=(x_1,y_2)$ belongs to  plane  curve $X'_1$ defined by non-algebraic equation of the form
\bq\label{s-curve}
X'_1:\qquad y^2-F(x)=0\,.
\eq
Second divisor $P_2=(x_2,y_2)$ lies on  horizontal line $X'_2$ defined by the  equation
\bq\label{h-line}
X'_2:\qquad y-k=0\,.
\eq
Thus,  we can formulate the following proposition.
\begin{prop}
At $a_1=a_2$ common level surface of the first integrals (\ref{ch-comm}) of  non-Hamil\-to\-nian vector field $Z$  (\ref{eq-m})
is a product of two plane curves $X'_1\times X'_2$ (\ref{s-curve}-\ref{h-line}).
\end{prop}
The proof is a straightforward calculation.

Divisors $P_1$ and $P_2$ determine compatible Poisson brackets,  recursion operators, algebra of Haantjes  operators with vanishing Nijenhuis torsion and  $2\times2$ Lax matrices.  For instance,
we can easily recover Poisson bivector $\mathcal P'$ (\ref{ch-poi2}) obtained by  brute force method in the previous Section
\[
\mathcal P'=f_1(x_1)\left(
 \dfrac{\partial}{\partial x_1} \wedge \dfrac{\partial}{\partial y_1}
-\dfrac{\partial}{\partial y_1}\wedge \dfrac{\partial }{\partial x_1}\right)
+f_2(x_2)\left(
 \dfrac{\partial}{\partial x_2} \wedge \dfrac{\partial}{\partial y_2}
-\dfrac{\partial}{\partial y_2}\wedge \dfrac{\partial }{\partial x_2}\right)\,,
\]
where $f_{1,2}$ are given by (\ref{ch-f12}).

Constructions of the Poisson brackets (\ref{poi-f}),  recursion operators (\ref{rec-f}) and algebra of Haantjes  operators
\[ L_0=Id\,,\qquad  L_i=
  \dfrac{\partial}{\partial x_i}\otimes d x_i
+\dfrac{\partial}{\partial y_i}\otimes d y_i,\quad i=1,2,
\]
 are independent from the form of plane curves  $X'_1$ and $X'_2$ and time variable, in contrast with  construction of the  $2\times2$ Lax matrices
\[
\mathcal L=\left(
    \begin{array}{cc}
      V & U \\
      W & -V\\
    \end{array}
  \right)\quad\mbox{and}\quad\mathcal A=\left(
                               \begin{array}{cc}
                                 0 & 1 \\
                                 S & 0 \\
                               \end{array}
                             \right)\,,\quad\mbox{so that}\quad \dot{\mathcal L} =[\mathcal L,\mathcal A]\,,
\]
here $x$ is a spectral parameter,  polynomials in $x$
\[
U(x)=x_2(x-x_1)\,,\qquad V(x)=\dfrac{1}{2}\,\dot{U}(x)\,,
\]
are the standard  Jacobi polynomials on a  product of plane curves $X'_1$ and $X'_2$ and
\[
W(x)=\dfrac{F(x)-V^2(x)}{U(x)}\,,\qquad S(x)=-\dfrac{\dot{W}(x)}{2V(x)}\,,
\]
are functions on $x$.

Unfortunately,  notion of the compatible Poisson brackets,  recursion operators, algebra of Haantjes  operators with vanishing Nijenhuis torsion and  Lax matrices can not help us in the search of real trajectories of motion.

\section{Conclusion}
In this note we discuss a non-Hamiltonian vector field  appearing in consideration of a  motion of the Chaplygin ball rolling on a horizontal plane which rotates with constant angular velocity  \cite{bbm18}.   In two partial cases we present division of this vector field by Hamiltonian vector fields using brute force computer calculations. In the first case one reduces equations of motion to Hamiltonian form which can also be  done in the framework of Hamiltonization  theory. In the second case vector field is a sum of two Hamiltonian vector fields which cannot be obtained by using any type of Hamiltonization. In both cases we obtain Turiel type deformations of canonical Poisson brackets on the cotangent bundle to the sphere.

 Calculation of at least two  different representations of a given vector field via Hamiltonian vector fields
 is a crucial part of finding  variables of separation in the Jacobi method.  In any existing  theory of Hamiltonization the main aim is to obtain one very special representation, which may not exist for the given mathematical model of a physical process,  and, therefore, these theories are not applicable to algorithmic  search of variables of separation. We prefer to develop a computer version of the Jacobi method which could be applicable both Hamiltonian and non-Hamiltonian vector fields with relatively low numbers of equations of motion. Of course, these computer algorithms  are not applicable to abstract nonholonimic systems with non-fixed arbitrary numbers of degrees of freedom.

We would like to thank A.V. Borisov and I.S. Mamaev for genuine interest and helpful discussions.

The work was supported by the Russian Science Foundation (project  15-12-20035).

\end{document}